\documentclass[aps,epsfig]{revtex4}
\usepackage{graphicx}
\begin{document}
\def\be{\begin{equation}}
\def\ee{\end{equation}}
\def\bfi{\begin{figure}}
\def\efi{\end{figure}}
\def\bea{\begin{eqnarray}}
\def\eea{\end{eqnarray}}

\title{Out of equilibrium dynamics of the spiral model}

\author{Federico Corberi}
\affiliation {Dipartimento di Matematica ed Informatica and 
INFN, Gruppo Collegato di Salerno, and CNISM, Unit\'a di Salerno,
Universit\`a di Salerno, 
via Ponte don Melillo, 84084 Fisciano (SA), Italy,
and Universit\'e Pierre et Marie Curie - Paris VI\\
Laboratoire de Physique Th\'eorique et Hautes Energies\\
4 Place Jussieu, 5\`eme \'etage, 75252 Paris Cedex 05, France}
\author{Leticia F. Cugliandolo}
\affiliation{Universit\'e Pierre et Marie Curie - Paris VI,
Laboratoire de Physique Th\'eorique et Hautes Energies,
4 Place Jussieu, 5\`eme \'etage, 75252 Paris Cedex 05, France}

\begin{abstract}
We study the relaxation of the bi-dimensional kinetically constrained
spiral model. We show that due to the reversibility of the dynamic
rules any unblocked state fully decorrelates in finite times
irrespectively of the system being in the unjammed or the jammed phase.
In consequence, the evolution of any unblocked configuration occurs in a
different sector of phase space from the one that includes the
equilibrium blocked equilibrium configurations at criticality and in
the jammed phase.  We argue that such out of equilibrium dynamics
share many points in common with coarsening in the one-dimensional
Ising model and we identify the coarsening structures that are,
basically, lines of vacancies.  We provide evidence for this claim by
analyzing the behaviour of several observables including the density
of particles and vacancies, the spatial correlation function, the
time-dependent persistence and the linear response.
\end{abstract}

\maketitle

PACS: 
05.70.Ln, 75.40.Gb, 05.40.-a

\section{Introduction} \label{intro}

Kinetically constrained models are toy
models for the glassy phenomenon~\cite{FA}
~(see~\cite{Jackle,Sollich,Leonard} for reviews). These models display
no thermodynamic singularity: their equilibrium measure is simply the
Boltzmann factor of independent variables and correlations only reflect
the hard core constraint. Still, they capture many features of real
glass-forming systems; among them one can mention stretched exponential
relaxation~\cite{FB}, super-Arrhenius equilibration time~\cite{GPG},
dynamical heterogeneity~\cite{KA} and self-diffusion/viscosity
decoupling~\cite{H}. They also display glassy out of equilibrium
aspects, such as physical aging~\cite{KPS} and effective
temperatures~\cite{S} (for a review of these aspects see \cite{Leonard}) that are
similar to the ones first obtained in random $p$-spin
systems~\cite{Cuku,Cukupe}, as well as heterogeneous
aging relaxation~\cite{Chamon}.  Bootstrap percolation arguments allowed
C. Toninelli {\it et al.} to show that most finite dimensional
kinetically constrained models studied so far do not even have a
dynamic transition at a particle density that is less than unity
in the thermodynamic limit~\cite{Cristina0}. On Bethe
lattices instead a dynamical transition similar to the one predicted
by the mode coupling theory might occur~\cite{Sellitto}.

A finite dimensional kinetically constrained model with an ideal {\it
  glass-jamming} dynamic transition at a particle density that is
different from one, the two-dimensional spiral model, was introduced
in a series of papers by C. Toninelli {\it et
  al.}~\cite{Cristina,Cristina1,Cristina2,Cristina3}. 
The dynamic transition is defined as the 
 critical density at which an {\it
  equilibrium} configuration can no longer be emptied by applying the
dynamic rule.  In consequence, at the transition an infinite cluster
of particles (or spins in
an equivalent representation) that are blocked by the dynamic 
constraints  exists. The model is defined on a square
lattice and the transition occurs at a value
of the control parameter $p$, the probability to update an unblocked site,  
that coincides with the critical
threshold of directed site percolation in $d=2$, that is to say,
$p_c\simeq 0.705$. The density of the frozen cluster is discontinuous
at $p_c$ which means that the frozen structure is compact rather than
fractal at criticality. Thus, at and above $p_c$ a finite fraction of
particles are blocked. It was shown in these papers that approaching
the critical $p_c$ from the unjammed phase the decorrelation of an
equilibrium configuration decays in two steps similarly to what is
common in glass-forming liquids. The time-scale for relaxation -- say
the $\alpha$ relaxation time -- diverges with a Vogel-Fulcher law
(faster than a power law). A crossover length below which finite size
effects are important also diverges faster than a power law when $p_c$ is
approached from below.

In this paper we study the dynamics of the $2d$ spiral model
numerically. For concreteness, we first show results for the
equilibrium dynamics, meaning the relaxation of an equilibrium
configuration, below but close to the critical $p_c$ -- the
`super-cooled liquid' -- and above $p_c$ -- the `equilibrium
glass'. Next, we focus on the out of equilibrium kinetics.
Specifically, we consider {\it annealing} and {\it quenching} processes. In the
former the system is prepared in an equilibrium initial condition at
$p_0\geq p_c$ and subsequently brought into the glassy regime with
$p>p_0$ by using a finite rate of change of the parameter $p$. In the
latter we study the evolution after a sudden {\it quench}, in which an
equilibrium state at $p_0=0$ (completely empty configuration) is
evolved from time $t=0$ onwards with the dynamic rule specified by a
different value of $p>p_0$.  This procedure is similar to the
temperature quench of a liquid.  For $p<p_c$, after a non equilibrium
transient the system attains the unblocked equilibrium state.  The
relaxation time for attaining such a state diverges as $p\to p_c$. In
the critical case with $p=p_c$, the system approaches the blocked
equilibrium state by means of an aging coarsening dynamics with some
points in common with the ones observed in critical quenches of
ferromagnetic models. Freezing is not observed because the dynamic
density $\rho (t)$ is always smaller than the critical one $\rho _c
=p_c$, at any finite time.  A different situation occurs for filling
at $p>p_c$.  Also in this case the system keeps evolving up to the
longest observed times and a blocked state is never observed despite
the fact that $\rho (t)$ exceeds $\rho _c$ for long enough times. This
can be understood by noticing that the dynamics are fully reversible:
for any $p$, each configuration reached dynamically can always evolve
back to the initial state with $\rho <p_c$ by the time reversed process,
although with a very low probability. Therefore, states with blocked
regions cannot be dynamically connected to the initial unblocked
states and they cannot be reached by any reversible process.  Frozen
regions being dynamically forbidden, the system approaches a high
density ($\rho > \rho_c$) ensemble which does not contain any of the
blocked states the measure of which is relevant in equilibrium. This
feature recalls what happens in coarsening systems where, again for
dynamical reasons, the kinetics occurs on the ensemble of
configurations with vanishing magnetization the weight of which is
negligible in equilibrium~\cite{Palmer}. Actually, it will be shown
that the dynamics of the spiral model at $p>p_c$ strongly resembles
coarsening in ferromagnets.

\section{Model} \label{model}

\begin{figure}
    \centering
   \rotatebox{0}{\resizebox{.5\textwidth}{!}{\includegraphics{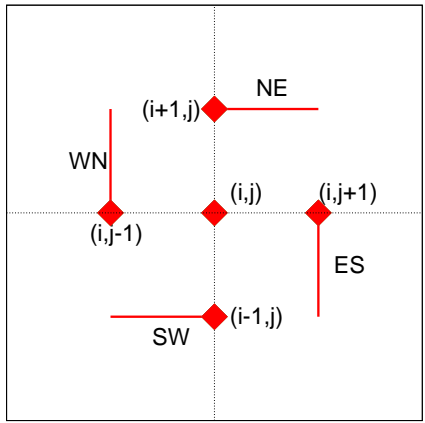}}}
\vspace{0.5cm}
    \caption{(Color online.) The neighbouring sites determining the
frozen or free to move character of the center site $(i,j)$.}
\label{figsketch}
\end{figure}

A binary variable $n_{ij} =0,1$ is defined on the sites $(i,j)$ of an
${\cal L}\times {\cal L}$ square lattice in $d=2$. $n_{ij}=1$
corresponds to a particle, $n_{ij}=0$ to a vacancy.  For a given
position $(i,j)$, let us define the following couple of neighbouring
sites (see Fig.~\ref{figsketch})
\begin{itemize}
\item{$(i,j+1)$,$(i+1,j+1)$, north east (NE) couple}
\item{$(i+1,j)$,$(i+1,j-1)$, south east (ES) couple}
\item{$(i,j-1)$,$(i-1,j-1)$, south west (SW) couple}
\item{$(i-1,j)$,$(i-1,j+1)$, north east (WN) couple}
\end{itemize}

The possibility to update the variable $n_{ij}$ depends on the
configuration of such couples. Specifically, if the sites belonging to
at least two consecutive couples (namely, NE+SE or SE+SW or SW+NW or
NW+NE) are empty (otherwise stated occupied by vacancies), then site
$(i,j)$ can be updated (either emptied or filled).  Otherwise it is
blocked.  We consider each site coupled to a particle reservoir in
such a way that particles can enter or leave the sample from its full
volume, in contrast to what was used in~\cite{KPS}, for instance,
where particles were allowed to access or leave the ($3d$ in this
case) system only through the borders.  We use periodic boundary
conditions.

Defining the variable $M_{ij}$ such that $M_{ij}=0$ if
site $(i,j)$ is frozen, and $M_{ij}=1$ otherwise the updating
rule can be compactly expressed in terms of the transition rates 
$W_p(n_{ij}\vert n'_{ij})$ to go from $n$ to $n'$ on site $(i,j)$ as
\begin{eqnarray}
W_p(0\vert 1)=M_{ij}\ p \qquad\qquad 
W_p(1\vert 0)=M_{ij}\ (1-p).
\label{transrate}
\end{eqnarray}

The equilibrium state has been studied in
detail by C. Toninelli {\it et al.}~\cite{Cristina,Cristina1}. 
The model can be regarded as a two-level system described by
the Hamiltonian $H[s]=-\sum _{ij}n_{ij}$ at the inverse temperature
\begin{equation}
\beta =\ln [p/(1-p)]
\; . 
\label{eq:beta}
\end{equation}
When $p\to 1/2$ the inverse temperature vanishes $\beta \to 0$ whereas
for $p\to 1$ it diverges $\beta \to \infty$.  The Bernoulli measure
implies $\rho=\langle n_{ij}\rangle =p$ for the equilibrium density of
particles.  This means that, in equilibrium at zero temperature ($p\to
1$), the lattice is full while at infinite temperature ($p\to 1/2$) it
is half-filled.  The existence of a blocked cluster at $p \geq p_c<1$
was shown by studying the critical $p$ above which an equilibrium
configuration cannot be emptied.

\begin{figure}
    \centering
   \rotatebox{0}{\resizebox{.3\textwidth}{!}{\includegraphics{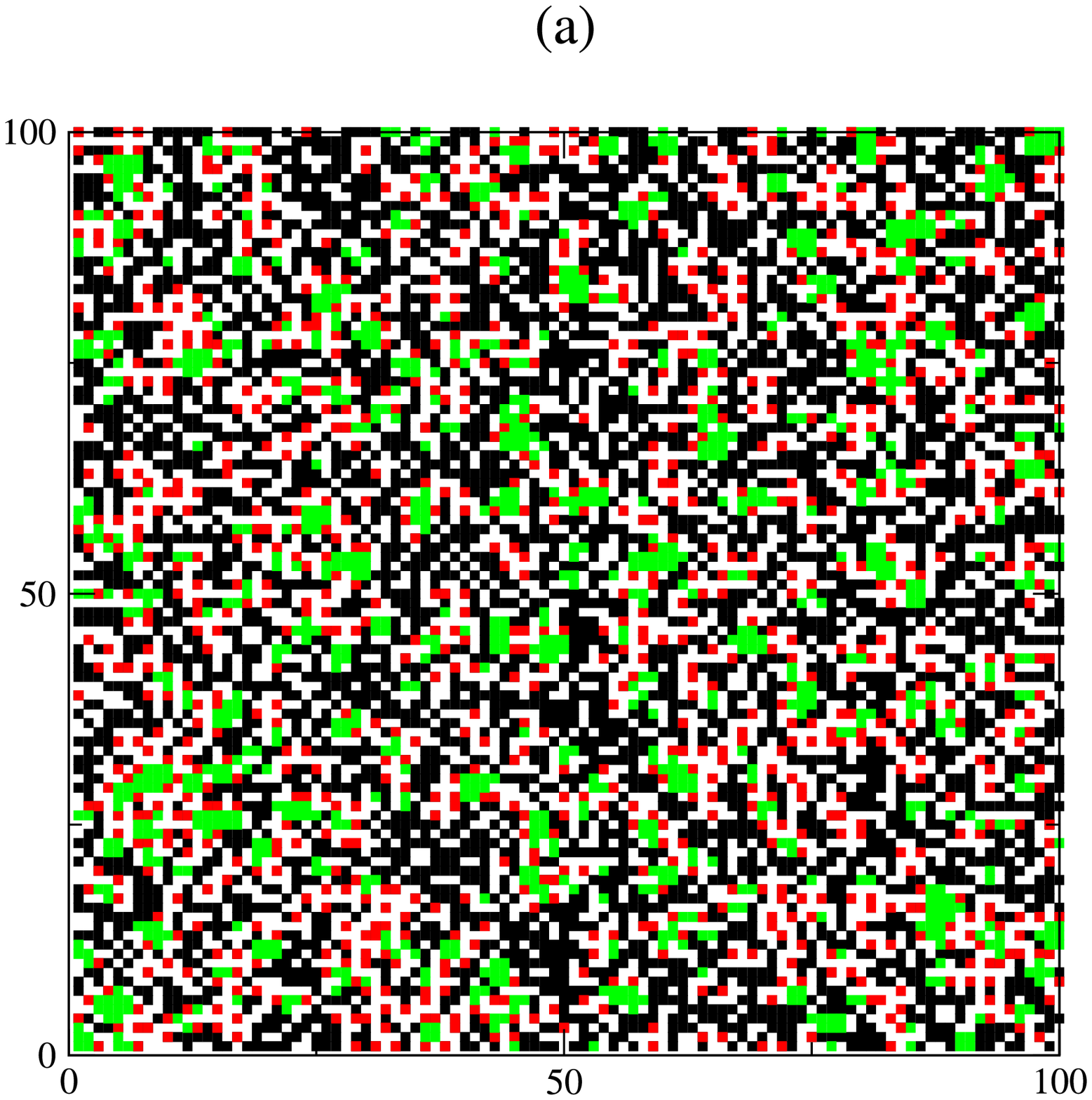}}}
\hspace{.5cm}
   \rotatebox{0}{\resizebox{.3\textwidth}{!}{\includegraphics{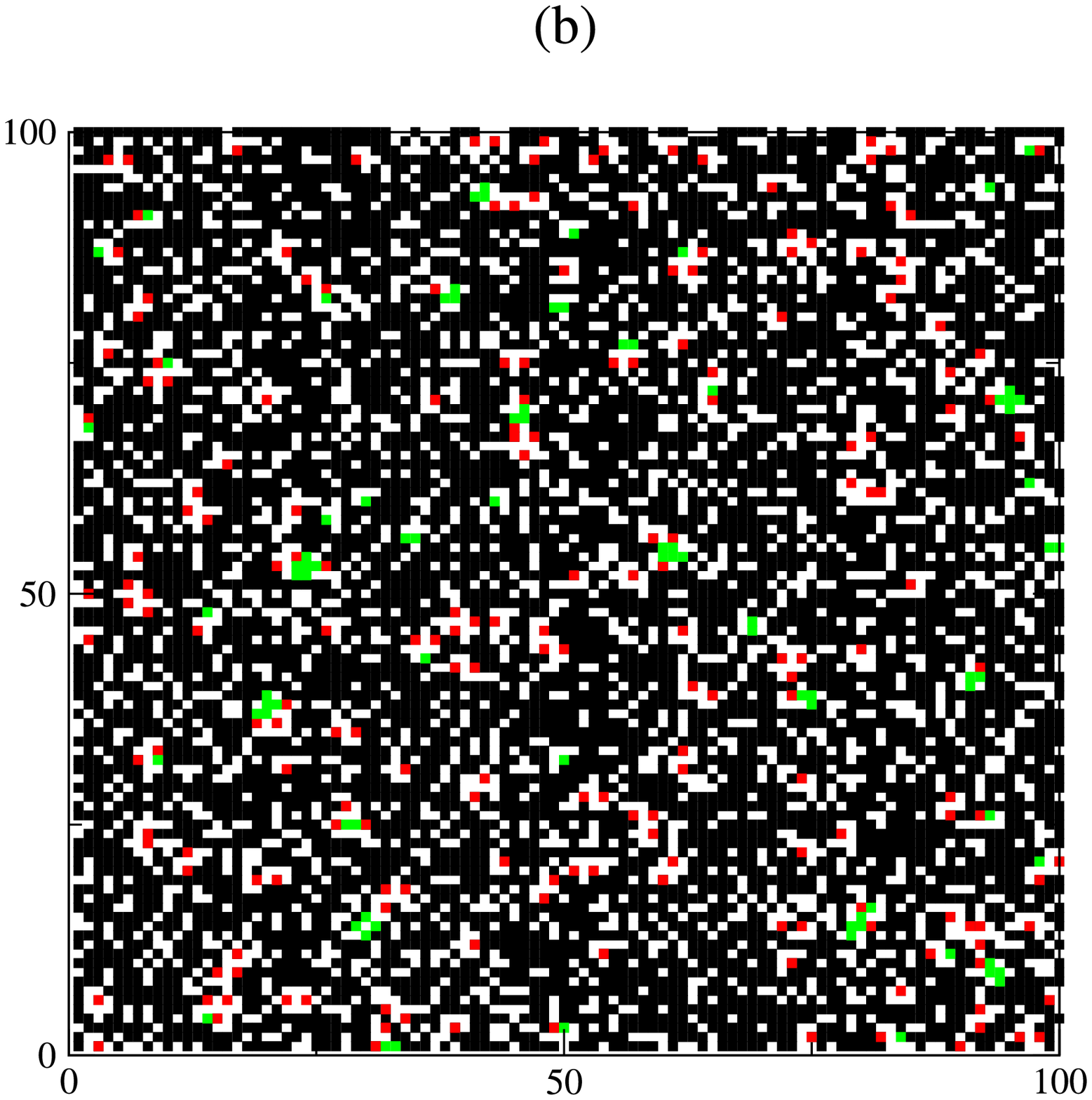}}}
\hspace{.5cm}
      \rotatebox{0}{\resizebox{.3\textwidth}{!}{\includegraphics{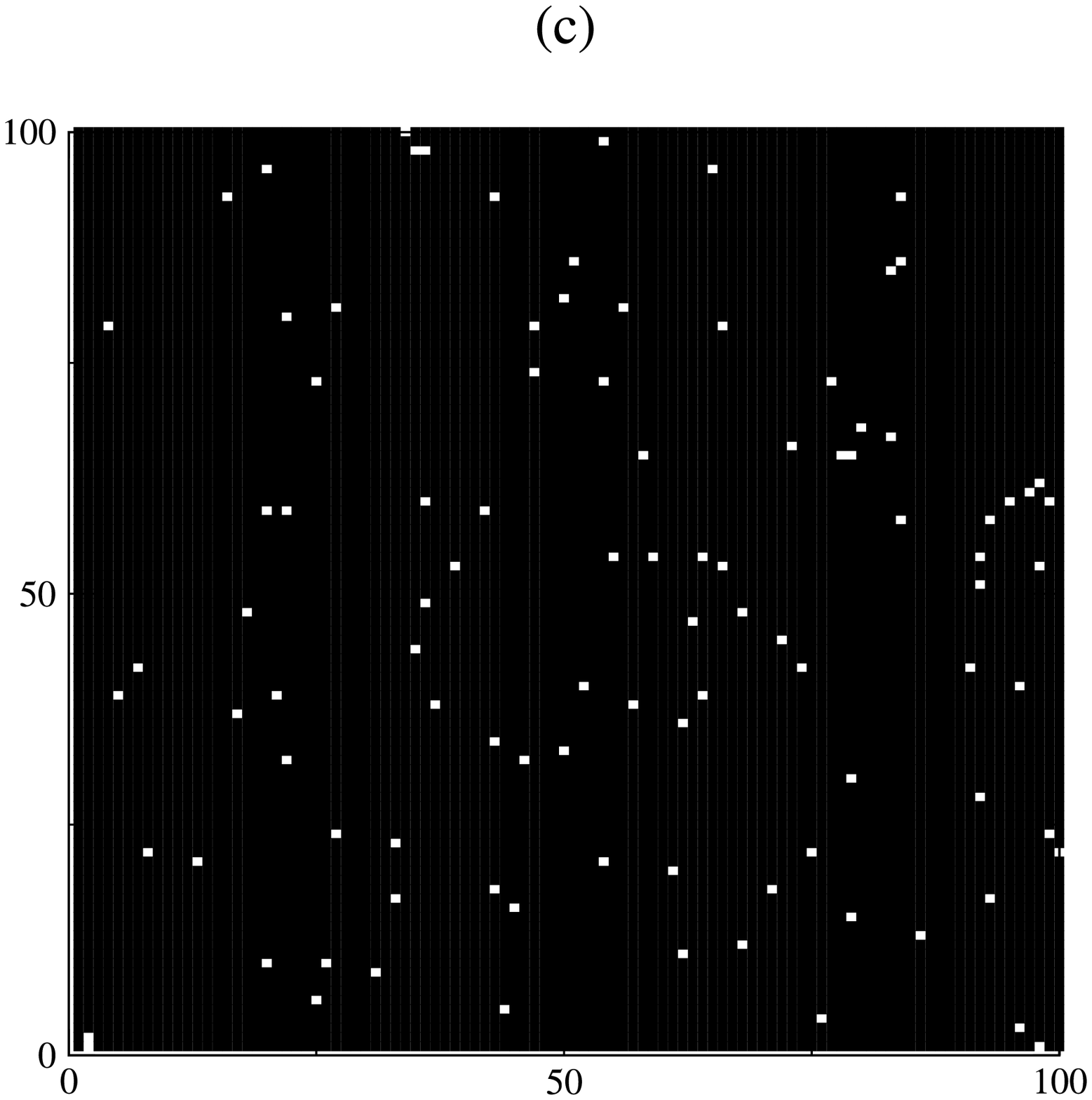}}}
\vspace{0.5cm}
\caption{(Colour online.) Equilibrium configurations at (a) $p=0.5<p_c$ ,
  (b) $p_c\simeq 0.705$, and (c) $p=0.99>p_c$. The colour code is such that: a
  filled blocked site is black; an empty blocked site is white; a free
  particle (one that can be removed) is red and a free vacancy (one
  that can be filled) is green. We show a small fraction of the sample to
  make the snapshot easier to visualize.}
\label{fig:conf-equil}
\end{figure}

Interestingly enough the microscopic dynamic rules are fully reversible. This implies that 
an initial configuration with no blocked structure can never reach, via the dynamics, a 
state with a blocked sub-ensemble of sites within. This fact holds for a system coupled to
the reservoir in any possible way (volume, border, or other). 
We shall elaborate on the consequences
of this fact later in the paper.

\section{Observables} \label{obs}

The dynamics of kinetically constraint models are usually
studied in terms of the probability that a spin never flips between
time $t_w$ and a later time $t$, also called the persistence
function $\phi(t,t_w)$. 
Operatively, we define this quantity as
\be
\phi(t,t_w)=\frac{1}{N}\langle N_{noup}(t,t_w)\rangle,  
\ee
where $N={\cal L} \times {\cal L}$ is the number of lattice points,
$N_{noup}(t,t_w)$ is the number of sites that have not been updated
between $t_w$ and $t$ in a given realization of the dynamics and  
$\langle \dots \rangle$
denotes an average over different histories and initial conditions.
In equilibrium conditions one has $\phi(t,t_w)=\phi(t-t_w)$. Moreover, 
at and above $p_c$ one has $\phi_\infty\equiv \lim_{t\to\infty}
\phi(t,t_w) >0$ due to the frozen backbone, while below $p_c$ the
asymptotic value vanishes $\phi_\infty =0$. 

Similar properties are exhibited by 
the autocorrelation function
\be
C(t,t_w)=\langle n_{ij}(t)n_{ij}(t_w)\rangle
-\langle n_{ij}(t)\rangle \langle n_{ij}(t_w)\rangle.
\label{autoc}
\ee

The impulsive autoresponse function is defined as
\be
R (t,t_w)= 
\left . \frac{\delta \langle n_{ij}(t)\rangle}{\delta \epsilon_{ij}(t')}\right \vert _{\epsilon_{ij}=0}
\label{r}
\ee
where $\epsilon _{ij}$ is a perturbation changing the transition rates
as $W_p(n_{ij},n'_{ij})\to W_{p+\epsilon _{ij}}(n_{ij},n'_{ij})$.
Notice that we dropped the site dependence in $C$ and $R$ due to space homogeneity. 
The integrated response function, or susceptibility, is
\be
\chi (t,t_w)=\int _{t_w} ^t dt' R(t,t')
\; .
\label{chi}
\ee 
In equilibrium conditions the response function is related to the
autocorrelation function by the fluctuation-dissipation theorem (FDT)
\be 
T\chi (t,t_w)=C(t,t)-C(t,t_w)
\label {fdt}
\ee 
where $T=\beta^{-1}$ with $\beta$ defined in Eq.~(\ref{eq:beta})
in the spiral model.
Out of equilibrium the response function can still be related to
correlation functions of the unperturbed system (with $\epsilon=0$)
according to the generalization of the FDT derived in~\cite{algo}.
This will allow us to determine the response function numerically
without applying any perturbation.  

It is useful to introduce the normalized
autocorrelation function 
\be 
\hat C(t,t_w)=C(t,t_w)/C(t,t) 
\ee 
and
susceptibility \be \hat \chi (t,t_w)=\chi (t,t_w)/C(t,t).  \ee

Finally, let us define
\be
l(t)=[\rho _{eq}-\rho (t)]^{-1}.
\label{pseudolungh}
\ee
In the non-equilibrium cases considered in Sec. \ref{quench}, when an initially
empty state is progressively filled with the transition rates $W_p$ in Eq.~(\ref{transrate}),
$\rho (t)$ always approaches $\rho _{eq}=p$. In this case $l(t)$ is a monotonically growing
function of time, and it will be convenient to re-parametrize $t$ in terms of $l$.

The spatial evolution of the structure is usually 
monitored via the equal time correlation function (equivalently, in Fourier 
space, the structure factor):
\be
G(r,t)=\langle n_{ij}(t)n_{i'j'}(t)\rangle-\langle n_{ij}(t)\rangle^2
\ee
where $r$ is the distance between sites $(i,j)$ and $(i',j')$.
A typical length can 
be extracted from $G$ by using
\be
L(t)=\left 
[\frac{\int dr \ r^\alpha \ G(r,t)}{\int dr \ G(r,t)}\right ]^{1/\alpha}.
\label{lungh}
\ee
When scaling holds, namely $G(r,t)/G(0,t)=g(r/L(t))$, changing $\alpha$ leads
to the same determination of $L(t)$ apart from a multiplicative constant.

All our simulations are done with square systems with linear size
${\cal L}=200-2000$, depending on the situation.
We typically average over $10^2-10^4$ realizations of the
dynamics. 

\section{Equilibrium relaxation} \label{equilibrium} 

Figure~\ref{fig:conf-equil} displays three panels with equilibrium configurations at
$p<p_c$ (a), $p=p_c$ (b) and $p>p_c$ (c). We shall later confront
these images to the ones reached dynamically from an empty state, see Fig.~\ref{figconfigs}.

\begin{figure}
    \centering
   \rotatebox{0}{\resizebox{.5\textwidth}{!}{\includegraphics{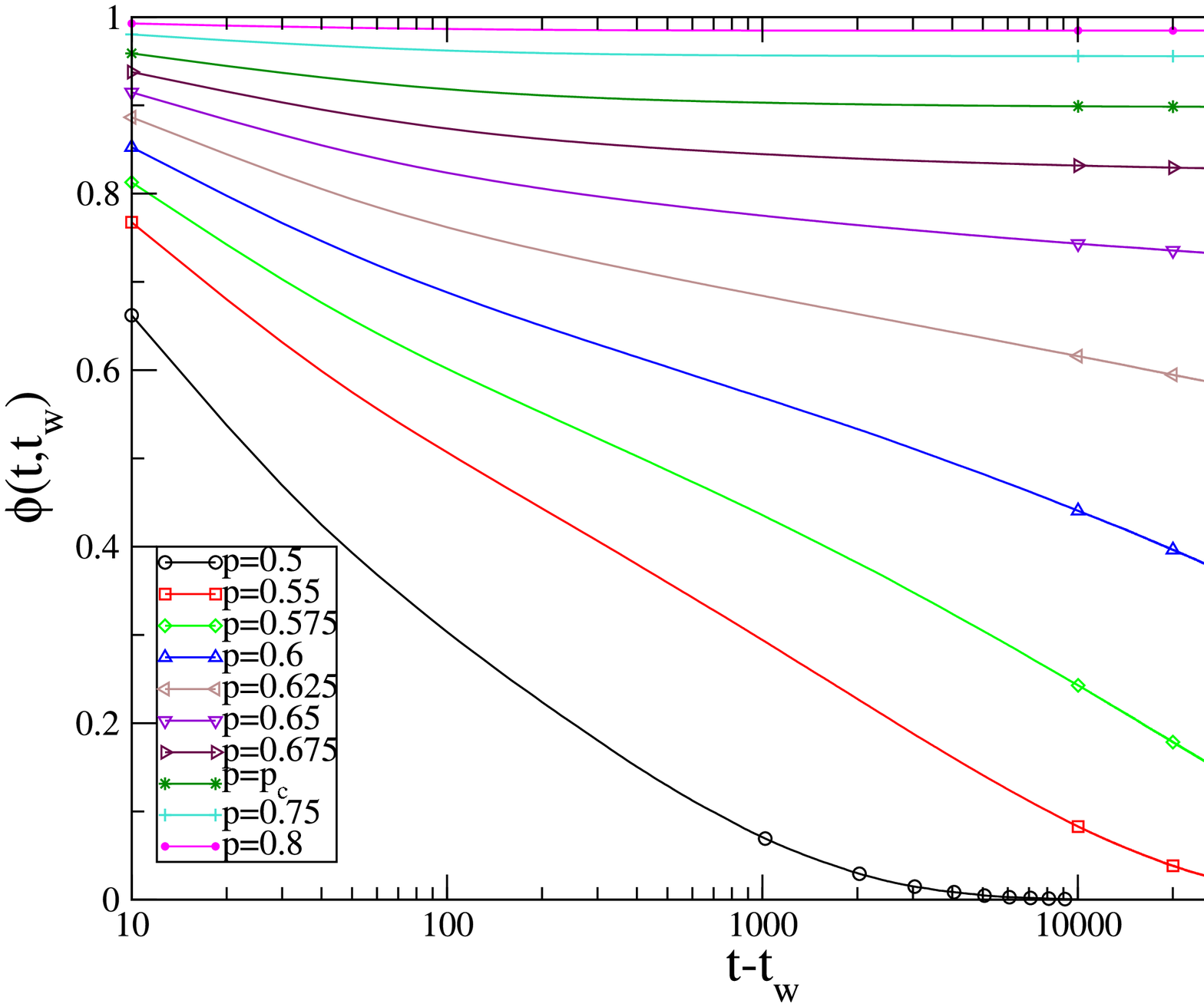}}}
   \vspace{1cm}

    \rotatebox{0}{\resizebox{.5\textwidth}{!}{\includegraphics{figpersistenceeqbis.eps}}}

   \vspace{0.5cm}

    \caption{(Color online.) Upper panel: Decay of the persistence function for an
    initial condition in equilibrium at different values of $p$.
    $\phi_\infty\equiv\lim_{t\to\infty}\phi(t)=0$ below $p_c$ and
    $\lim_{t\to\infty}\phi(t)>0$ above $p_c$.  Lower panel: Dependence
    of $\phi_\infty$ on $p$. In the inset $\phi _\infty (p)-\phi
    _\infty (p_c)$ is plotted against $p-p_c$ in a double logarithmic
    scale, for $p>p_c$. The red dashed line is the best fit to the
    power-law behavior $\phi _\infty (p)-\phi _\infty (p_c)\propto
    (p-p_c)^\alpha$, yielding $\alpha =0.901$.}
\label{fig:persistence-eq}
\end{figure}

In Fig.~\ref{fig:persistence-eq}(a) we study the relaxation of an equilibrium configuration
for several values of $p$ by means of the persistence function. 
For values of $p<p_c$ there is no frozen structure in the 
system and the persistence function, $\phi$,  decays to zero. Close to the critical 
$p_c$, although a blocked spanning structure does not exist, 
long living quasi-frozen structures do. The decay of the persistence 
or the normalized correlation close (but below) criticality is reminiscent
of the hallmark of the super-cooled liquid relaxation that is associated
to the formation of long-lived {\it cages} surrounding each particle~\cite{Cavagna}.
The persistence of these structures diverges as $p_c$ is approached.
As a consequence, $\phi (t,t_w)$ develops a plateau that becomes longer
and longer when $p$ gets closer to $p_c$ as in a super-cooled liquid. This numerical 
result confirms the analytic ones in~\cite{Cristina,Cristina1,Cristina2,Cristina3}.

Above $p_c$ the initial configurations have a finite
fraction of frozen particles so $\phi$ decays to a constant,
$\phi_\infty$, that depends on $p$. 
Figure~\ref{fig:persistence-eq}(b)
shows $\phi_\infty(p)$. As expected
$\phi_\infty$ is different from zero at $p_c$; we found
$\phi_\infty(p_c) \simeq 0.9$, a relatively large value. A fit to the
data close to the critical point yields $\phi_\infty(p) \simeq
\phi_\infty(p_c) + c (p-p_c)^\alpha$ with $\alpha \simeq 0.9$.
$\phi_\infty$ increases with $p$
and reaches $1$ at $p\to 1$.

The normalized autocorrelation function $\hat C(t,t_w)$ behaves in a qualitatively
similar way.
Moreover, consistently with equilibrium dynamics  $\hat C(t,t_w)$ is
related to $\hat \chi (t,t_w)$ 
by the FDT, namely $T\hat \chi (t,t_w)=1-\hat C(t,t_w)$, see Eq.~(\ref{fdt}). 

\section{Out of equilibrium relaxation} \label{results}

In this Section we discuss the out of equilibrium dynamics of a system
initially prepared in some equilibrium configuration at $p_0$ and then
evolved from time $t=0$ onward with the transition rates
(\ref{transrate}) and $p>p_0$.  As discussed in Sec. \ref{model} this
 procedure is the analogue of  a cooling experiment in thermal systems.

 We shall consider two types of `cooling' procedures: the former is a
 slow process where a system at equilibrium at $p=p_0\geq p_c$, a
 configuration with a blocked structure, is evolved with the
 transition rates (\ref{transrate}) and a time-dependent $p=p_0+rt$,
 $r$ being the equivalent of a cooling rate.  The latter is an
 instantaneous quench where an equilibrium system at $p_0=0$, a
 completely empty state, is evolved using $p<p_c$, $p=p_c$ and $p>p_c$
 as target values of the control parameter for the subsequent
 dynamics.

We recall that the dynamic rule is reversible and configurations with
or without a blocked sub-ensemble of sites are mutually inaccessible
even for finite systems, a fact that draws an important difference
with usual stochastic models without dynamical constraints.

\subsection{Slow cooling from equilibrium at $p_c$: threshold level}
\label{slowcooling}

\begin{figure}
    \centering
   \rotatebox{0}{\resizebox{.5\textwidth}{!}{\includegraphics{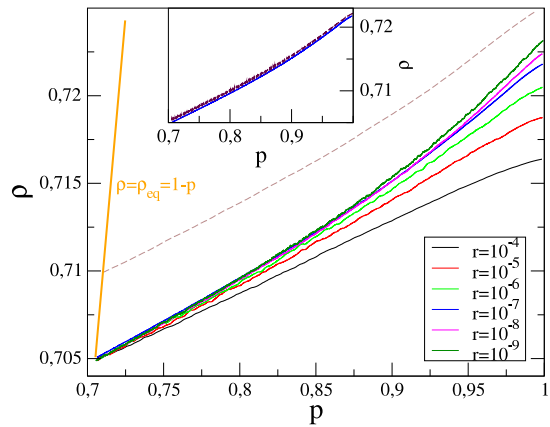}}}
\vspace{0.5cm}
\caption{(Color online.) The density of particles as a function of $p$ for linear
  cooling rate procedures $p(t)=p_0+rt$ starting from equilibrium at
  $p_0 \ge p_c$. Continuous curves: $p_0=p_c\simeq 0.705$ and
  different cooling rates $r$ given in the key.  The annealed
  densities at fixed $p$ increase with decreasing $r$ but they remain
  well below the equilibrium values, $\rho_{eq}=p$, shown with an
  orange straight line in the figure.  Dashed line: the initial
  condition is equilibrium at $p_0=0.71>p_c$ and the cooling rate is
  $r=10^{-7}$. In the inset the continuous and the dashed line
  represent the curves obtained with $r=10^{-7}$ starting from
  $p_0=p_c$ (continuous line) or $p_0=0.7051>p_c$.}
\label{threshold}
\end{figure}

In Fig.~\ref{threshold} we show the density of particles found after
cooling with a finite rate. Continuous lines with
different colour represent processes starting from an equilibrium
configuration at $p_c$ and using different cooling rates $r$. Dashed
curves, instead, refer to cases where cooling starts from equilibrium
above $p_c$.  Concerning cooling from equilibrium at $p_c$, for
$p$'s that are above but not too far from $p_c$ the curves for
$r<10^{-7}$, say, superpose. For larger values of $p$ there is a
residual $r$ dependence that presumably disappears for even
smaller $r$'s. We assume then that the dark green curve is quite close
to the one for the limit of vanishing cooling rate $\rho_{th}(p)
\equiv \lim_{r\to 0} \rho_r(p)$.  $\rho_{th}(p)$ is definitely below
the equilibrium density $\rho_{eq}$ for all $p>p_c$ and hits a limit $\rho_{th}(1)
\gtrsim 0.723< 1$ when $p\to 1$.

This analysis demonstrates that the blocked structure at $p_c$
involves vacancies that cannot be filled when increasing $p$ with the
cooling procedure and the dynamic density remains well below the
equilibrium one (shown with a continuous straight orange line in
Fig.~\ref{threshold}).  A similar procedure in which $T$ is changed infinitely 
slowly,  carried out in the
$p$-spin model with $p\geq 3$ defines the {\it threshold
free-energy level}~\cite{Cuku} that remains higher than the
equilibrium free-energy in the full low temperature phase.

Cooling with a finite rate an initial condition equilibrated
at a value $p_0(>p_c)$ yields other levels of blocked states (low lying
metastable TAP states in the $p\geq 3$-spin analogy). In Fig~\ref{threshold} 
we show with a dashed line one of such curves with $p_0=0.71$. 
A more careful analysis shown in the inset suggests that the curves 
for different $p_0$ do not cross in the full $p>p_c$ region, again in 
accordance with the $p$-spin scenario in which there is neither level crossing 
nor level merging when $T$ is lowered infinitesimally slowly.

\subsection{Infinitely rapid quench from equilibrium at $p_0=0$.} \label{quench}

In this Section we consider the dynamical process whereby an empty
configuration is filled with the transition rate (\ref{transrate}),
for different values of $p$. We have checked that starting from
an equilibrium configuration corresponding to $0<p_0<p_c$ (i.e. not
completely empty), does not change the evolution significatively.
In the case of quenches to $p>p_c$ the effect of $p_0>0$ is
to delay the asymptotic regime where, as we shall discuss below, scaling holds.
This is due to the more blocked nature of the initial state.
For this reason, the choiche $p_0=0$ is the most convenient
to study the asymptotic properties, and we shall always make this choice
in the following. 
 
\begin{figure}
    \centering 
   \rotatebox{0}{\resizebox{.5\textwidth}{!}{\includegraphics{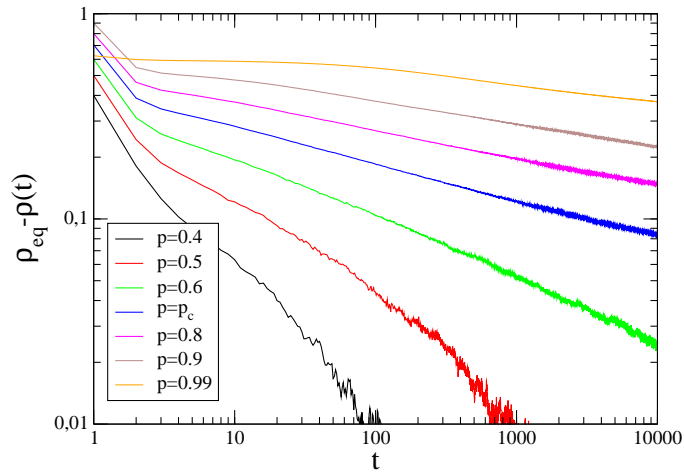}}}
\vspace{0.5cm}
    \caption{(Color online.) The difference between $\rho (t)$ and its equilibrium
value $\rho _{eq}$ against time for different values of
$p$. The relaxation is faster the smaller the value of $p$.}
\label{figrho-quench}
\end{figure}

In Fig.~\ref{figrho-quench} we plot the
difference between $\rho _{eq}$ and the dynamic density $\rho (t)$.
This quantity is related to $l(t)$ through Eq.~(\ref{pseudolungh}).
For $p<p_c$ the density decays exponentially to the equilibrium value
over a typical time that increases as $p$ approaches $p_c$.  At
$p=p_c$, there is a power law decay $\rho (t)-\rho _{eq}\propto
t^{-\nu (p_c)}$ with $\nu(p_c)\simeq 0.16$.  For $p>p_c$, the same
behavior, $\rho (t)-\rho_{eq}\propto t^{-\nu (p)}$ is observed,
although the exponent $\nu (p)$ becomes smaller as $p$ increases and
seems to go to zero in the limit $p\to 1$.  Notice that this implies
that $l(t)$ grows as a power law in the whole phase $p\geq p_c$, and
logarithmically in the extreme limit $p\to 1$. The relaxation seems to
be smooth in the sense that $\rho(t)$ decays as a power law within the
whole explored time-window.

If one extrapolates the power law behaviour to infinitely long times
the fact that $\rho_{eq}-\rho(t) \to 0$ would mean that, surprisingly
enough, the relaxation avoids the threshold level -- defined by the
procedure described in Sect.~\ref{slowcooling}.  Part of the
explanation of this fact is that the dynamics are reversible and after
a quench the system never reaches configurations with a completely blocked
sub-ensemble. This implies that the system cannot go through threshold
states to lower lying metastable ones since all the former do have
blocked sub-ensembles. This is clearly different from what is the
$p$-spin scenario in which a typical initial condition as the one
corresponding to infinite temperature first relaxes to the threshold
level~\cite{Cuku,Cavagna,Montanari} and then penetrates below it via
thermal activation for finite size systems. It is also different from
the behaviour of kinetically constrained spin models on the Bethe
lattice~\cite{Sellitto} that seem to conform to the $p$-spin scenario.
The fact that the out of equilibrium relaxation of the spiral model
reaches $\rho(t)>\rho_{th}(p)$ implies that the dynamics should go to
the long-time configurations `turning around' blocked ones in phase
space.

In the following Sections we shall discuss the three cases of quenches to $p<p_c$,
$p=p_c$ and $p>p_c$ in more detail separately.

\subsubsection{Quench from $p_0=0$ to $p<p_c$: equilibrium dynamics.}
 
\begin{figure}
    \centering 
   \rotatebox{0}{\resizebox{.5\textwidth}{!}{\includegraphics{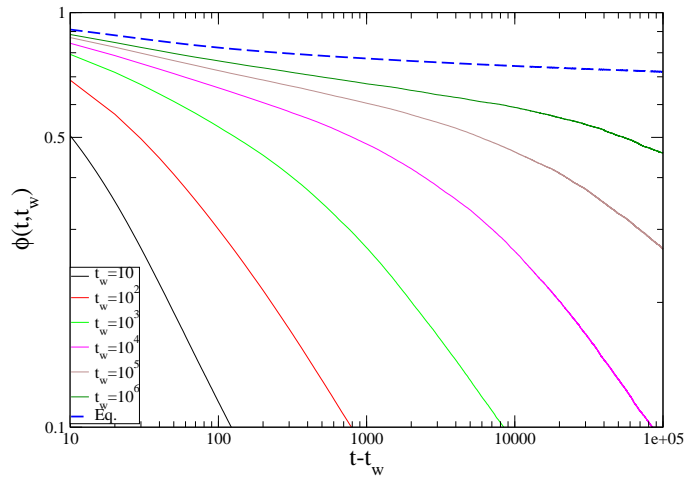}}}
\vspace{0.5cm}
\caption{(Color online.) The persistence function $\phi (t,t_w)$ is plotted against
  $t-t_w$ in the case of a quench from an empty configuration to
  $p=0.65$, for different values of $t_w$. The equilibrium curve (long
  dashed) is approached in the large-$t_w$ limit. The system is never
  blocked and $\phi$ decreases to zero in all cases.}
\label{figphinoneqsotto}
\end{figure}

We have already observed that for $p<p_c$ but sufficiently close to
$p_c$ long lasting quasi-blocked structures exist which are
responsible for the development of a plateau in the persistence
function. In the case considered here, since the initial configuration
is empty, those structures must be built. Hence the plateau observed
in equilibrium (Fig.~\ref{fig:persistence-eq}) can be observed only in
the limit $t_w\to \infty$, when equilibrium is attained and the long
lived quasi-frozen structures are fully developed. For finite $t_w$,
on the other hand, those structures are not completed and the plateau
has a shorter extent, as shown in Fig. \ref{figphinoneqsotto}.  $\hat
C(t,t_w)$ behaves similarly.

\subsubsection{Quench from $p=0$ to $p=p_c$: Critical quench.}

\begin{figure}[h]
\vspace{1cm}
    \centering 
   \rotatebox{0}{\resizebox{.5\textwidth}{!}{\includegraphics{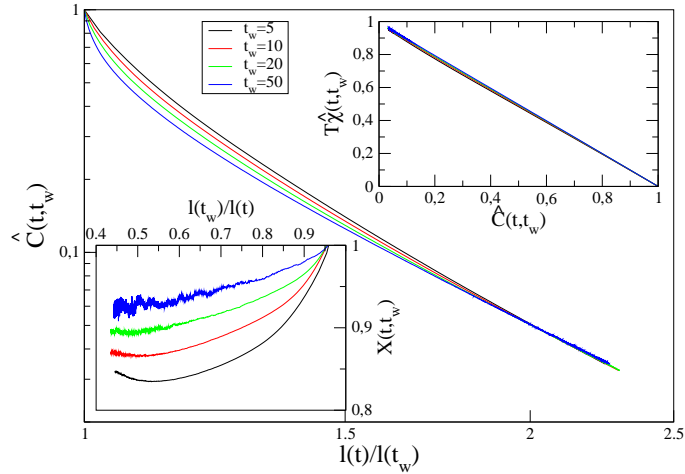}}}
\vspace{0.5cm}
\caption{(Color online.) Dynamics after a quench from $p_0<p_c$ to $p_c$.  $\hat C
  (t,t_w)$ is plotted against $l(t)/l(t_w)$. The upper inset shows the
  fluctuation-dissipation plot of $\hat \chi (t,t_w)$ against $\hat C
  (t,t_w)$.  The lower inset shows the fluctuation-dissipation ratio
  $X(t,t_w)$ as a function of $l(t_w)/l(t)$.}
\label{figscalingcrit}
\end{figure}

In this case a configuration with a blocked structure is approached in an infinite time
but never reached.
Since, as shown in Fig.~\ref{figrho-quench}, $\rho (t)$ approaches the equilibrium value
asymptotically as a power law, we shall re-parametrize $t$ in terms of $l(t)$ in the
following. We do this because, although $t$ and $l$ are trivially related
in the asymptotic time domain, at early times when corrections to the power law 
behavior of $l$ are present, this quantity turns out to be better suited to
enlighten the scaling properties of the dynamics.
Since $\phi (t,t_w)$ and $\hat C(t,t_w)$ display the same scaling properties
we discuss in the following the behavior of the latter quantity.
The data of Fig.~\ref{figscalingcrit} show that $\hat C(t,t_w)$
obeys the scaling 
\be
\hat C(t,t_w)=f\left [\frac{l(t)}{l(t_w)}\right ].
\label{scalc}
\ee
The collapse of the curves for different $t_w$ is very good for large values
of $x=l(t)/l(t_w)\stackrel{>}{\sim} 1.75$, while it is poorer for smaller values of $x$ perhaps
due to preasymptotic effects. 
The same scaling is found for the susceptibility, as testified by the fact that
the parametric plot of $\hat \chi (t,t_w)$ against $\hat C(t,t_w)$
(upper inset of Fig. \ref{figscalingcrit}) does not depend on $t_w$.
The parametric plot
has the linear equilibrium behavior $T_c\hat \chi (t,t_w)=1-\hat C(t,t_w)$, as in
ferromagnetic models quenched to $T_c$.
However, while in the latter case the quantity $X(t,t_w)=TR(t,t_w)/(\partial C/\partial t_w)$ is a 
non trivial function of $t/t_w$ with the limiting value 
$\lim _{t_w\to \infty } \lim _{t\to \infty }X(t,t_w)=X_\infty <1$~\cite{xinfty},
here $X(t,t_w)$ slowly approaches $X(t,t_w)\equiv 1$ in the limit $t_w\to \infty $,
as shown in the lower inset of Fig.~\ref{figscalingcrit}.

\subsubsection{Quench from $p_0=0$ to $p>p_c$: Coarsening.} \label{coarsening}

We now discuss the dynamics following an infinitely rapid quench from 
equilibrium at $p_0=0$ to $p>p_c$. We argue that the relaxation 
occurs through a coarsening process which resembles some features of
ferromagnetic systems. The kinetics simplifies in the limit $p\to 1$ because
$W_1(1\vert 0)/W_1(0\vert 1) \to 0$ and the annihilation of particles
can be neglected with respect to creation events if the latter are available.
We shall consider this case first, since a cleaner description
of the kinetics and of its basic mechanisms is possible. The
case with a non-vanishing $1-p$ finite, where the finite annihilation
probability adds new mechanisms to the kinetics, will be considered
further below.

\vspace{0.5cm}
\noindent
{\it Quenches to} $p\to 1$
\vspace{0.5cm}

\begin{figure}
    \centering
   \rotatebox{0}{\resizebox{.3\textwidth}{!}{\includegraphics{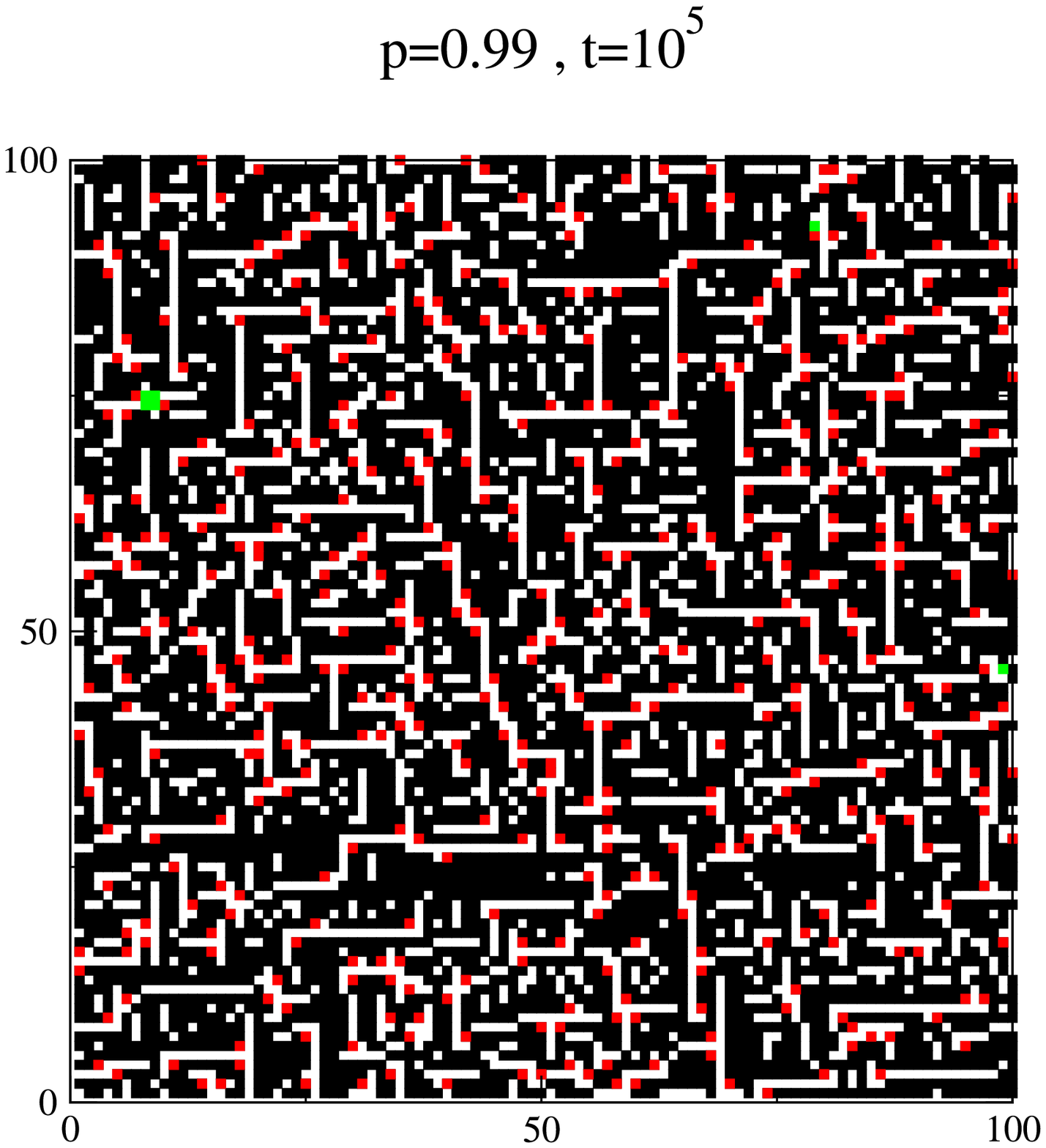}}}
\hspace{.5cm}
   \rotatebox{0}{\resizebox{.3\textwidth}{!}{\includegraphics{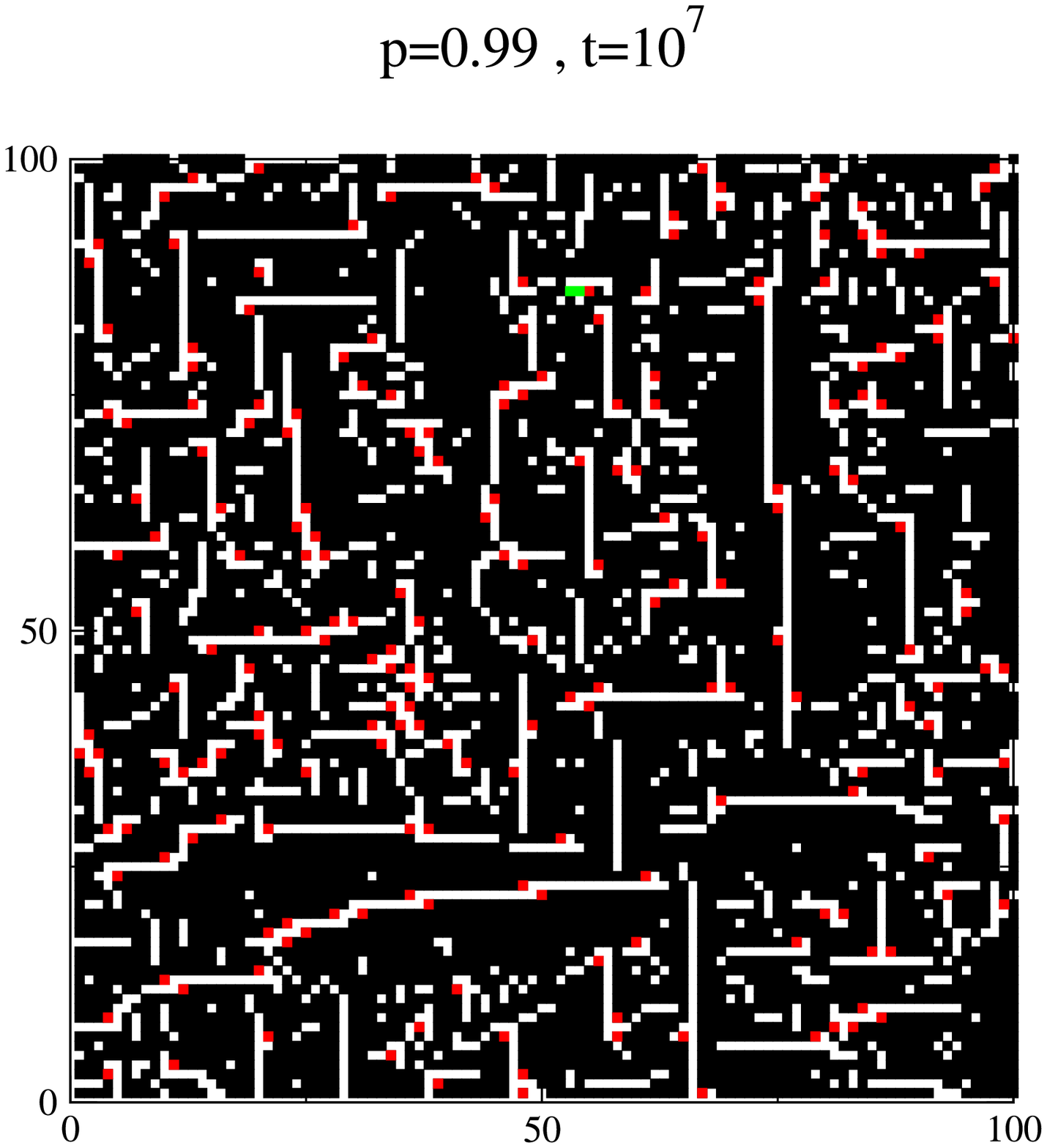}}}
\hspace{.5cm}
   \rotatebox{0}{\resizebox{.3\textwidth}{!}{\includegraphics{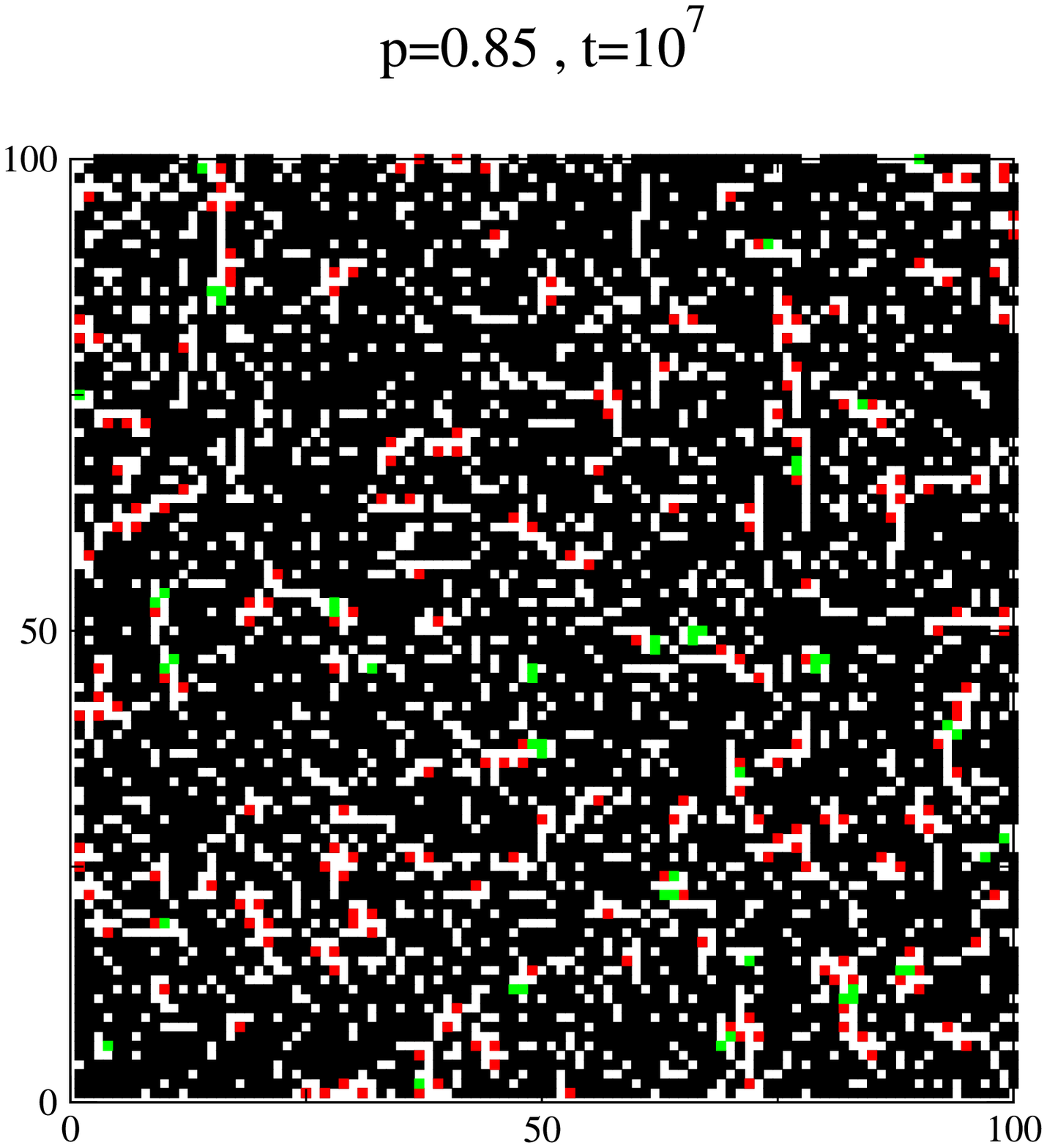}}}
\vspace{0.5cm}
\caption{(Color online.) Left and central panel: A zoom on the configuration of the
  system at two times ($t=10^5$ and $t=10^7$) after a quench to
  $p=0.99$.  Black and white sites are frozen particles and vacancies,
  respectively.  Red and green sites are particles and holes that can
  be updated.  Right panel: Configuration of the system at time
  $t=10^7$ after a quench to $p=0.85$.  The structure is fuzzier than
  in the $p=0.99$ case.}
\label{figconfigs}
\end{figure}

\begin{figure}[h]
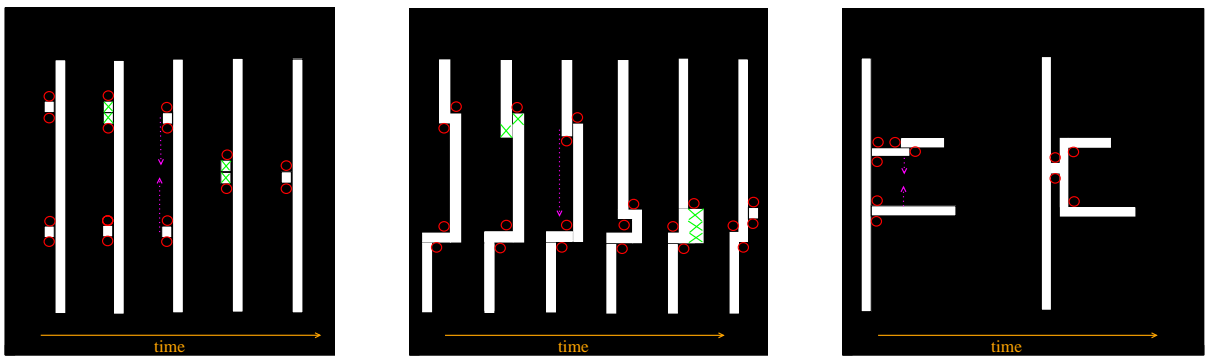

\vspace{1cm}
    \centering
   \rotatebox{0}{\resizebox{.26\textwidth}{!}{\includegraphics{figmonomers.eps}}}
   \hspace{.5cm}
   \rotatebox{0}{\resizebox{.27\textwidth}{!}{\includegraphics{figkink.eps}}}
   \hspace{.7cm}
   \rotatebox{0}{\resizebox{.27\textwidth}{!}{\includegraphics{figtjunk.eps}}}
\vspace{0.5cm}
    \caption{(Color online.) Schematic evolution of the system. Black sites are particles,
    (a red circle indicates the ones which are not frozen) white sites are vacancies
    (a green cross indicates those which are not frozen). Left panel: the mechanism
    producing the diffusion and annihilation of two TVs. Central panel: the mechanism
    producing the diffusion and coalescence of two kinks. Right panel: two
    T-junctions can diffuse similarly to kinks (see central panel) and merge.}
\label{figschematic}
\end{figure}

In this section we study the $p\to 1$ limit by letting $p=0.99$
in the numerical simulations.
In the limit $p\to 1$, $\rho (t)$ still decays to zero, see Fig.~\ref{figrho-quench}. 
Hence, for large times there is a very small fraction
of vacancies. Despite this fact, vacancies cannot be isolated, otherwise
the dynamical constraints would freeze the system, which is not
observed at any time in our simulations. These two requirements
can be both fulfilled if vacancies segregate into 
quasi-unidimensional domains (see Fig.~\ref{figconfigs}).
A representation of a quasi-unidimensional domain is given in Fig.~\ref{figschematic}.
Such a domain is basically a set of vacancies
each of which is surrounded by $n=1,2$ neighbouring vacancies.
The word {\it quasi} refers to the fact that these strings of vacancies can be
seldom decorated by some bi-dimensional feature such as a 
{\it traveling vacancy} (TV), and $T$-junctions,
as shown in Fig.~\ref{figschematic}. Since these objects are
basically one-dimensional the system is allowed to build very
long strings with a vanishing vacancy density in the thermodynamic 
limit.   

In the limit $p\to 1$ the dynamics proceed as follows:
starting from a generic configuration at large times the only 
possible move is almost always the removal of a non frozen particle.
This process takes a typical time 
$\tau _0\simeq (1-p)^{-1}$, which diverges as $p\to 1$. There is
basically no evolution on time-scales shorter than $\tau _0$ which
must be regarded, therefore, as the shortest time in the system.
Whenever a particle is removed it is always possible
to add a particle (in the site from which it was previously removed
or in the neighbourhood) and this event almost always occurs
for $p\to 1$. On the other hand, processes such as the consecutive
removal of two (or more) particles, which have probability
of order $(1-p)^2$ (or higher powers of $1-p$) can be discarded. 
Therefore we can focus on the leading events to lowest order in 
$1-p$ and the dynamics are provided by the following mechanisms
(see Fig.~\ref{figschematic})
\begin{itemize}
\item{Diffusion and annihilation of traveling vacancies: with
probability $1-p$ one of the particles surrounding a traveling vacancy
can be removed.  When this occurs a pair of adjacent TVs is formed and,
since $p\simeq 1$, one of the two is immediately filled with a
particle. This process, which taken alone does not change $\rho $, makes TVs
diffuse along flat parts of the string. If two TVs come together one of
them is immediately filled with a particle and $\rho$ decreases.}
\item{Deformation around kinks or T-junctions: a particle can be
removed with probability $1-p$
from the internal part of a kink or a T-junction. 
Hence the neighbouring site can be filled by a particle.
This process produces the deformation of the string, since kinks and
T-junctions can diffuse and/or generate new kinks and TVs.  In
addition, if two kinks or T-junctions meet, a new particle can be
added lowering the density.  Notice that these mechanisms, besides
lowering the density, may give rise to the coalescence of strings,
increasing the size of the quasi-unidimensional
aggregates of vacancies, a phenomenon which is clearly observed in
Fig.~\ref{figconfigs}}.

\end{itemize}

As discussed above the deformation and coalescence between
strings lead to a perpetual growth of the typical string length,
and consequently of the average distance between strings $L(t)$, 
as can be observed  in Fig.~\ref{figconfigs}.
This suggests that dynamical scaling may set in 
for large times, due to the prevalence of $L(t)$ over
any other length, similarly to what happens in coarsening systems. 
According to the scaling hypothesis, configurations of the
system look statistically similar if distances are measured
in units of $L(t)$. For the equal time correlation function
this implies $G(r,t)/G(0,t)=g(r/L(t))$. Similarly, the autocorrelation function 
must depend on $t,t_w$ only through the ratio $x=L(t)/L(t_w)$,
namely
\be
\hat C(t,t_w)=f(x)
\qquad \mbox{with} \qquad
x=\frac{L(t)}{L(t_w)}
\ee
In the following we check this hypothesis.

\begin{figure}
    \centering
   \rotatebox{0}{\resizebox{.5\textwidth}{!}{\includegraphics{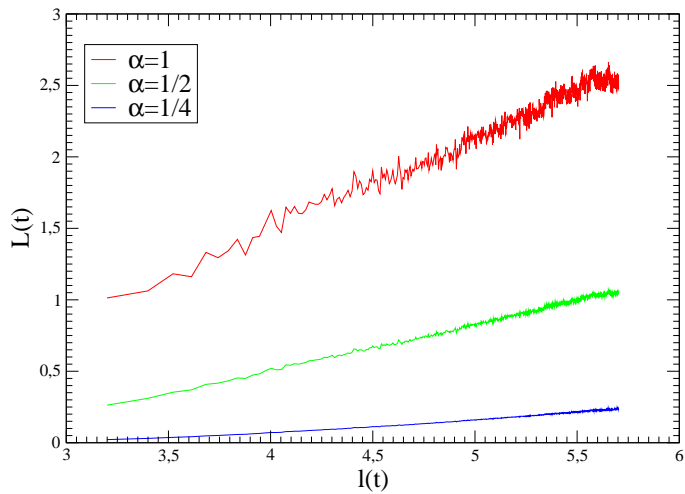}}}
\vspace{0.5cm}
    \caption{(Color online.) The typical length defined in Eq.~(\ref{lungh}), for different values
    of $\alpha $ given in the key, against $l(t)$.}
\label{figlengthvsrho}
\end{figure}

Due to the one-dimensional geometry of the strings,
we expect the typical distance between them to be proportional 
to the inverse of the vacancy density $\rho _v(t)$. Since in this limit
$\rho _v(t)=1-\rho (t)\simeq l^{-1}(t)$ we claim   
\be
L(t)\propto l(t). 
\label{lvsrho}
\ee
This is demonstrated
in Fig.~\ref{figlengthvsrho} where we plot $L(t)$ computed through
Eq.~(\ref{lungh}) for different values of $\alpha $ against $l(t)$. 
In the presence of 
scaling, as already observed in Sec. \ref{obs}, determinations
of $L(t)$ using different values of $\alpha $
give the same result, apart from a multiplicative constant. 
Notice however that, in the presence of noise, the quality of the 
determination of $L(t)$ may strongly depend on the value of
$\alpha $. Actually, different values of $\alpha $ weight differently
regions with different $r$ and, since data at small $r$ are typically less
noisy, smaller values of $\alpha $ lead to cleaner results. 
This is clearly observed in Fig. \ref{figlengthvsrho}.  
On the other hand, if preasymptotic corrections to scaling are present,
deviations from scaling can be enhanced or suppressed 
by changing $\alpha$, because those corrections may be
more effective in the large or small $r$ regions, depending on the system.
Actually, the data for $\alpha =1$ and $\alpha =0.5$ show no appreciable
deviations from the asymptotic law $L(t)\propto l(t)$ in the whole 
time range; with $\alpha =0.25$ one observes small deviations up to
$l(t)\simeq 4-4.5$ which can be interpreted as due to preasymptotic
corrections to scaling. In any case, data for $l(t)>4.5$ clearly show
that $L(t)\propto l(t)$ for all $\alpha $. This implies that scaling is obeyed
and that, in this case, $l(t)$ has the meaning of a length.  

\begin{figure}[h]
    \centering
\vspace{1cm}
   \rotatebox{0}{\resizebox{.5\textwidth}{!}{\includegraphics{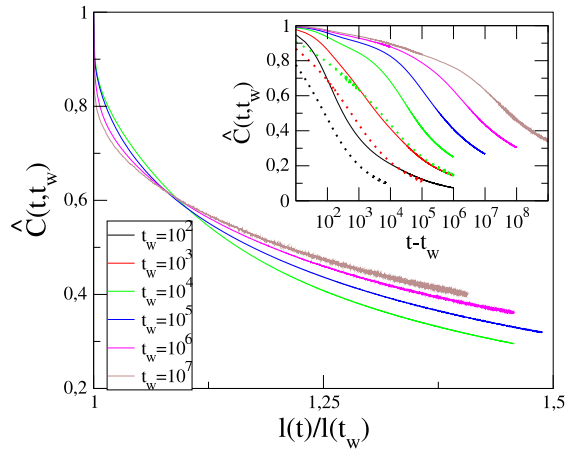}}}

\vspace{1cm}

\caption{(Color online.) The system is prepared in equilibrium at $p_0=0$ and then
  instantaneously quenched to $p=0.99$ at $t=0$.  $\hat C(t,t_w)$ is
  plotted against $l (t)/l (t_w)$ for different values of $t_w$ given
  in the key.  In the inset, the continuous lines are the same curves
  $\hat C(t,t_w)$ of the main figure but plotted against $t-t_w$. The
  dotted lines represent $\hat C(t,t_w)$ against $t-t_w$ (for
  different values of $t_w$ given by the same colour code of the
  continuous lines) after a quench to $p=0.85$.}
\label{figscalingc}
\end{figure}

Since $G(r,t)$ is too noisy to obtain further evidence for scaling we
consider in the following two-time quantities.  In the inset of
Fig.~\ref{figscalingc} we plot $\hat C(t,t_w)$ against $t-t_w$ for
different values of $t_w$. One clearly observes an aging behavior. The
behaviour is very similar to the one in the $1d$ Ising chain. The apparent
plateau slowly moves upwards and tends to $1$.  In the main figure
$\hat C(t,t_w)$ is plotted against $l (t)/l (t_w)$.  Although the data
collapse does not provide a definitive evidence for scaling, there is
a clear improvement upon the collapse as $t_w $ increases, suggesting
that scaling may be asymptotically obeyed for very large $t_w$.
Interestingly all the curves intersect at the point of coordinate $x
\simeq 1.08$ with curves for longer (shorter) $t_w$ lying below (above) curves for 
shorter (longer) $t_w$ before (after) the crossing point.
This is somewhat similar to what was found in Monte Carlo simulations of 
the Sherrington-Kirkpatrick (SK) spin-glass~\cite{Yoshino}, that is known to 
satisfy dynamic ultrametricity asymptotically~\cite{Cuku94}. However, in the spiral model 
the distance between consecutive curves reduces as $t_w$ increases
(contrary to what is seen in the SK case) thus suggesting that this 
peculiar fact is just a pre-asymptotic effect.

\begin{figure}[h]
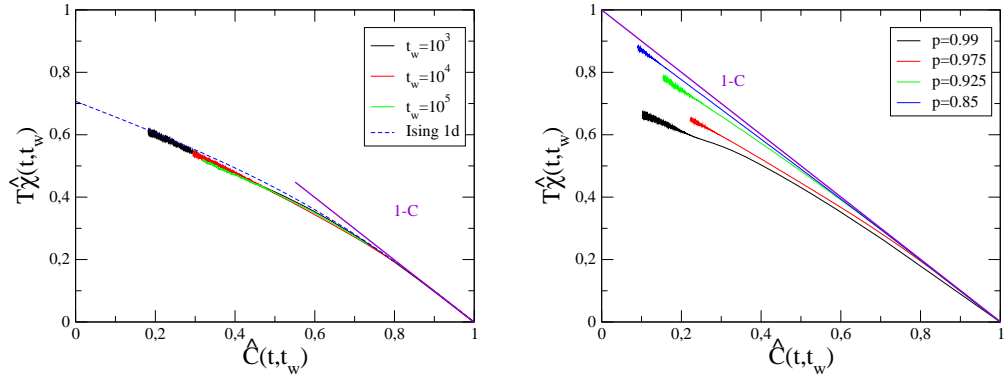

\vspace{1cm}
    \centering
   \rotatebox{0}{\resizebox{.35\textwidth}{!}{\includegraphics{figchidic.eps}}}
\hspace{0.5cm}
   \rotatebox{0}{\resizebox{.35\textwidth}{!}{\includegraphics{figchidic2.eps}}}

\vspace{1.5cm}

\caption{(Color online.) Left panel: $T\hat \chi(t,t_w)$ against $\hat C(t,t_w)$ for
  different values of $t_w$, in the case of an infinitely rapid quench
  from equilibrium at $p_0=0$ to $p=0.99$.  The bold purple line is
  the equilibrium behavior $T\hat \chi(t,t_w)=1-\hat C(t,t_w)$.  The
  dashed blue line is the parametric plot of $\chi $ vs $C$ in the
  $1d$ Ising model. Right panel: $T\hat \chi(t,t_w)$ against $\hat
  C(t,t_w)$ for $t_w=10^3$ in the case of an infinitely rapid quench
  from equilibrium at $p_0=0$ to different values of $p>p_c$ given in
  the key.}
\label{figchidic}
\end{figure}

Next we consider the response function. In Fig.~\ref{figchidic} the
fluctuation dissipation plot
$T\hat \chi (t,t_w)$ against $\hat C(t,t_w)$ is shown.  The curves collapse quite well on a
mastercurve $\widetilde \chi _{p\simeq 1}(\hat C)$.  This implies that
the response function obeys the same scaling as $\hat C(t,t_w)$,
namely $\hat \chi (t,t_w)=h(x)$ with $x=L(t)/L(t_w)$.  In the short time
regime, for values of the $x$ axis that are close to one, one has $\tilde \chi
(\hat C)_{p\simeq 1}\simeq 1-\hat C$.  As $\hat C$ decreases the curve
bends.  Although, naively, one could think that the parametric plot is
one with two straight lines (one with the temperature of the bath and
another one with a different value with a sharp crossover in between)
the interpretation of the dynamics in terms of one dimensional
coarsening suggests instead that the asymptotic construction is a
non-trivial curve analogue to the one found in the $1d$ Ising
model. Actually, as shown in Fig. \ref{figchidic}, $\widetilde
\chi_{p\simeq 1}(\hat C)$ is quite similar to the parametric plot
$\widetilde \chi _{T\simeq 0} (C)$ (known analytically~\cite{1dIM}) of
$\chi $ vs $C$ in the $1d$ Ising model quenched to $T\simeq 0$ (no normalization 
is needed in this case since $C(t,t)=1$).  Since
$\widetilde \chi _{T\simeq 0} (C)$ is known not to be universal
(depending, for instance on the observable used to compute the
response~\cite{sollich2}) we do not expect the curves for the two models
to be quantitatively equal, but only to behave in a qualitatively
similar way.  On the other hand, since the limiting slope $X_\infty=d
\widetilde \chi _{T\simeq 0}(C)/dC \vert _{C=0}=1/2$ is expected to be
universal~\cite{xinfty,Sollich}, it would be interesting to compare this
value with the limiting slope of the spiral model.

\vspace{0.5cm}
\noindent
{\it Quenches to} $p_c<p< 1$
\vspace{0.5cm}

In this Section we consider quenches from an empty state to $p_c<p<1$.
With respect to the case $p\to 1$ considered above, there is now 
a non-vanishing annihilation probability, due to a finite $1-p$. Hence, after
a time of order $\tau _n\simeq (1-p)^ {-n}$, processes where $n$
vacancies are created start to be observed.  Then, for $t<\tau _2$ the
dynamics proceeds similarly to the case of a quench to $p\to 1$; later
on a bubble of empty sites can be created around a quasi-dimensional
structure, as can be seen in the right panel of
Fig. \ref{figconfigs}.  These processes produce a local rejuvenation
of the system, since in the bubbles the density decreases (although
globally it is still growing as shown in Fig. \ref {figrho-quench}).
As an effect, for a given $t_w$, the autocorrelation function decays
faster than in the case with $p\to 1$, as can be seen in the inset of
Fig. \ref{figscalingc}.

The process of bubble formation can be compared to the fast quasi-equilibrium processes 
occurring in ferromagnetic systems when quenched to a finite $T>0$. In scalar models 
(e.g. the Ising model) these processes are basically the reversal of spins in the bulk of 
ordered domains due to thermal fluctuations. In systems with a finite critical temperature
$T_c$ (say, the Ising model in $d>1$) quenched below $T_c$,
 in the large-$t_w$ limit the timescale of fast processes is widely separated
from the timescale over which irreversible or aging events occur. 
Quasi-equilibrium processes are responsible for 
the fast initial decay of $C(t,t_w)$ to a plateau value $C(t,t_w)=m^2$,
namely the squared equilibrium magnetization,
which occurs for $t-t_w<\tau _{eq}$,  
where $\tau _{eq}$ is the equilibrium relaxation time.
Being equilibrium fluctuations in nature, these processes obey the FDT,
and Eq.~(\ref{fdt}) is observed in a restricted time domain when $C(t,t_w)\geq m^2$.
Despite this fact, 
the system is globally aging, as shown by the further decay of $C(t,t_w)$ from the 
plateau value on much larger timescales $t-t_w\gtrsim t_w \gg \tau _{eq}$
when aging processes become relevant. 
Notice that, by raising $T$ towards $T_c$
the region of the parametric plot where the FDT 
is observed extends to lower and
lower values of $C$ until at $T_c$ Eq.~(\ref{fdt}) 
is obeyed in the whole range of $C$ variation.
For system with $T_c=0$ (i.e. the $1d$ Ising model) 
quenched to a low $T>0$, instead, the appearance of thermal fluctuations in the bulk 
of domains at a certain time $t_{eq}$ is accompanied by the global equilibration of the system,
so that aging is interrupted.
Eq.~(\ref{fdt}) is then globally obeyed from $t_w \simeq t_{eq}$ onwards. 

In the spiral model with $p<1$ bubble formation seems to play a role broadly 
similar to thermal fluctuations in ferromagnets, although combining features
of systems with $T_c=0$ to those of the cases with $T_c>0$, and with some notable
differences. Actually,
when bubbles begin to be formed $\hat C(t,t_w)$ starts to decay faster than in the case with 
$p\to 1$, but we could not detect the existence  of a two step relaxation (with a plateau in between) 
in our data for $\hat C(t,t_w)$ (see Fig.~\ref{figscalingc}). 
This suggests that an effective time-scale separation between fast 
and slow processes as observed in ferromagnetic systems is not present in the spiral model,
at least in the time sector studied in our simulations. 
In Fig.~\ref{figchidic}-right the parametric plot of $\hat \chi$ against $\hat C$ is shown for
quenches to different values of $p>p_c$. Only the value $t_w=10^3$ is shown, we found that
for each choice of $p$ the curves with values of $t_w$ up to $t_w=10^5$ are  
indistinguishable within numerical errors from the corresponding ones with $t_w=10^3$.
Although we cannot exclude that a tiny $t_w$-dependence could be observed by considering 
much larger values of $t_w$, this strongly suggests that the curves with different $t_w$
collapse on a $p$-dependent mastercurve $\widetilde \chi_{p}(\hat C)$. 
Notice that by lowering $p$ towards $p_c$, $\widetilde \chi_{p}(\hat C)$ rises
towards the FDT line (\ref{fdt}) and the region of large $\hat C$ where the FDT
is approximatively obeyed is enlarged. This is partly similar to what observed
in ferromagnets with $T_c>0$ quenched below $T_c$ when $T$ is raised towards $T_c$. 
However, at variance with those systems, the shape of the plot is non trivial
(i.e. all the curves bend continuously) and there is no apparent dependence on $t_w$,
although from the inspection of $C$ one concludes that the system 
is still globally aging. 
These features resemble what observed in ferromagnets with $T_c=0$; in the spiral
model however this behavior is observed in the whole glassy phase $p>p_c$.
    
\section{Conclusions}

At variance with most kinetically constrained spin models, the spiral model
is characterized by an equilibrium glass-jamming transition 
at a finite value of 
the control parameter: for $p\geq p_c$ an infinite cluster of 
frozen particles exists. 
In view of the more general issue of glassy systems, therefore,
it is of a certain interest to
study and understand the effect and the relevance of such a transition on the non-equilibrium
dynamical properties.
In so  doing we uncover a quite interesting and rich scenario with
unexpected far from equilibrium features.
Actually, while the equilibrium dynamics (and the near-to-equilibrium one
of the annealing process considered in Sec.~\ref{slowcooling}) 
feels the presence of blocked regions the size of which diverges at
the transition, leading to diverging relaxation times in the
persistence function or in the autocorrelation, 
the non-equilibrium kinetics following an abrupt
change of $p$, from below to above $p_c$, takes a completely different
route avoiding the percolative blocked configurations relevant
in equilibrium. At the basis of this behavior is the reversibility
property of any kinetics obeying detailed balance. Due to this fact,
a blocked configuration cannot be connected dynamically with an
unblocked one. As a consequence, a system prepared initially in
any state without a spanning frozen cluster keeps evolving on a restricted
ensemble of configurations without such clusters. Evolution
on a sub-ensemble of configurations (those with vanishing magnetization)
is a key feature of coarsening systems. Actually, we found that
the analogy between the spiral model and quenched ferromagnets 
goes beyond this. The non-equilibrium kinetics of the spiral model
is actually a coarsening phenomenon where quasi-unidimensional
strings of vacancies deform, diffuse and merge increasing
their typical length $L(t)$ in a power-law fashion (logarithmically
in the extreme case $p\to 1$). Observable quantities 
obey a dynamical scaling symmetry with $L(t)$ acting as a scale factor,
much in the same way as in usual coarsening. Interestingly enough,
the non-equilibrium fluctuation dissipation plot is a non trivial
($p$-dependent) curve, qualitatively similar to the one observed in 
the $1d$ Ising model.
This feature may suggest that a similar mechanism is at work
in the kinetics of the spiral model and of the Ising chain.
It must be stressed, however, that in the former 
a non-trivial fluctuation-dissipation plot is observed
in the whole glassy phase with $p>p_c$, and not only in a limiting
sector of the phase diagram (i.e. $T\to 0$) as in the $1d$ Ising
model, a feature resembling what is found in the Sherrington-Kirkpatrick
spin-glass model~\cite{Cuku94}. 
    
\vspace{1cm}
\noindent
\underline{Acknowlewdgements.} We thank G. Biroli and Y. Shokef  
for very useful discussions. F.C. acknowledges financial support
from PRIN 2007 JHLPEZ ({\it Statistical Physics of Strongly correlated
systems in Equilibrium and out of Equilibrium: Exact Results and 
Field Theory methods) and from CNRS and thanks the LPTHE Jussieu
for hospitality during the preparation of this work.
L.F. Cugliandolo is a member of Institut Universitaire de France.


\begin{thebibliography}{99}
\bibitem{FA} G. H. Fredrickson and H. C. Andersen, Phys. Rev. Lett. {\bf 53}, 1244 (1984). 

\bibitem{Jackle} J. J\"ackle, J. Phys. Cond. Matt. 
{\bf 14}, 1423 (2002). 

\bibitem{Sollich} P. Sollich and F. Ritort, Adv. in Phys. 
{\bf 52}, 219 (2003). 

\bibitem{Leonard}
For a review see S. Leonard, P. Mayer, P. Sollich, L. Berthier, and J. P. Garrahan,
J. Stat. Mech. P07017 (2007). 

\bibitem{FB} G. H. Fredrickson  and S. A. Brawer, J. Chem. Phys. {\bf 84}, 3351 (1986).

\bibitem{GPG} I. S. Graham, L. Pich\'e, and M. Grant, Phys. Rev. E {\bf 55}, 2132 (1997).

\bibitem{KA} W. Kob and H. C. Andersen, Phys. Rev. E {\bf 48}, 4364 (1993).
J. P. Garrahan and D. Chandler, Phys. Rev. Lett. 
A. C. Pan, J. P. Garrahan, and D. Chandler, Phys. Rev. E {\bf 72}, 041106 (2005).

\bibitem{H} P. Harrowell, Phys. Rev. E {\bf 48}, 4359 (1993).
L. Berthier, D. Chandler, and J. P. Garrahan, Europhys. Lett. {\bf 69}, 320 (2005).
A. C. Pan, J. P. Garrahan, and D. Chandler, 
 Chem. Phys. Chem. 6, 1783 (2005).

\bibitem{KPS} 
G. H. Fredrickson and S. A. Brawer, J. Chem. Phys. {\bf 84}, 3351 (1986).
G. H. Fredrickson, Annals NY Acad. of Sci. {\bf 484}, 185, (1986).
J. Kurchan, L. Peliti, and M. Sellitto M, Europhys. Lett. {\bf 39}, 365  (1997).

\bibitem{S} M. Sellitto, 
Euro. Phys. J. B {\bf 4}, 135 (1998). R. L. Jack, L. Berthier, and J. P. Garrahan,
J. Stat. Mech. P12005 (2006). 

\bibitem{Cuku} L. F. Cugliandolo and J. Kurchan, Phys. Rev. Lett. {\bf 71}, 173 (1993). 

\bibitem{Cukupe} L. F. Cugliandolo, J. Kurchan, and L. Peliti, 
Phys. Rev. E {\bf 55}, 3898 (1997).

\bibitem{Chamon} 
C. Chamon, P. Charbonneau, L. F. Cugliandolo, D. R. Reichman, and M. Sellitto, 
J. Chem. Phys. {\bf 121}, 10120 (2004).


\bibitem{Cristina0} C. Toninelli, G. Biroli, and D. S. Fisher, Phys. Rev. Lett. 
{\bf 92}, 185504 (2004). 

\bibitem{Sellitto} 
J. Reiter, F. Mauch and J. J\"ackle, Physica A {\bf 184}, 458 (1992).
S. J. Pitts, T. Young, and H. C. Andersen, J. Chem. Phys. {\bf 113}, 8671 (2000).
M. Sellitto, G. Biroli, and C. Toninelli, Europhys. Lett. {\bf 69}, 496 (2005).  

\bibitem{Cristina}
G. Biroli and C. Toninelli, 
Eur. Phys. J. B {\bf 64}, 567 (2008).

\bibitem{Cristina1}
C. Toninelli and G. Biroli,
J. Stat. Phys. {\bf 130}, 83-112 (2008).

\bibitem{Cristina2}
C. Toninelli, G. Biroli, and D. S.  Fisher,
Phys. Rev. Lett. {\bf 96} 035702 (2006).

\bibitem{Cristina3}
C. Toninelli, G. Biroli, and D. S. Fisher,
Phys. Rev. Lett. {\bf 98} 129602 (2008).

\bibitem{Palmer} R.G. Palmer, Adv. Phys. {\bf 31}, 669 (1982).
J. Kurchan and J. Laloux, J. Phys. A {\bf 29}, 1929 (1996).
C.M. Newman and D.L. Stein, J. Stat. Phys. {\bf 94}, 709 (1999). 

\bibitem{Szamel} 
G. Szamel, J. Chem. Phys. {\bf 121}, 3355 (2004). 

\bibitem{algo} 
E. Lippiello, F. Corberi and M. Zannetti, Phys. Rev. E {\bf 71}, 036104 (2005). 

\bibitem{Cavagna} A. Cavagna,
{\it 
Supercooled Liquids for Pedestrians}  
 arXiv:0903.4264, Phys. Rep. (to appear). 

\bibitem{Montanari} A. Montanari and F. Ricci-Tersenghi, Phys. Rev. B 70, 134406 (2004). 

\bibitem{1dIM} C. Godr\`eche and J-M Luck, 
E. Lippiello and M. Zanetti, 

\bibitem{TAP} S. Franz and G. Parisi, J. Physique I {\bf 5}, 1401
  (1995).  A. Barrat, R. Burioni, and M. M\'ezard, J. Phys. A {\bf
  29}, L81 (1996).  L. F. Cugliandolo, J. Kurchan, P. Le Doussal, and
  L. Peliti, Phys. Rev. Lett. {\bf 78}, 350 (1997).

\bibitem{xinfty} C. Godr\`eche and J-M Luck, J. Phys.: Cond. Matt. 
{\bf 14}, 1589 (2002).  

\bibitem{sollich2} P. Sollich, S. Fielding, and P. Mayer, 
J. Phys.: Cond. Matt. {\bf 14}, 1683 (2002). 
P. Mayer, L. Berthier, J.P. Garrahan, and P. Sollich,
Phys. Rev. E {\bf 68}, 016116 (2003).

\bibitem{Yoshino} H. Takayama,  H. Yoshino, and  K. Hukushima
J. Phys. A {\bf 30}, 3891 (1997).

\bibitem{Cuku94} L. F. Cugliandolo and J. Kurchan, J. Phys. A 
{\bf 27}, 5749 (1994).

\end{thebibliography}
\end{document}